# Artificial Intelligence-Based Techniques for Emerging Robotics Communication: A Survey and Future Perspectives

S. H. Alsamhi[1], Ou Ma[2], M. S. Ansari[3]

[1]School of Aerospace, Tsinghua University, Beijing, China, and IBB University, Ibb, Yemen
[2]College of Engineering and Applied Science, University of Cincinnati, Ohio, USA
[3]Department of Electronics Engineering, AMU, Aligarh, India

salsamhi@tsinghua.edu.cn, ou.ma@uc.edu, samar.ansari@zhcet.ac.in.

**Abstract:-** This paper reviews the current development of artificial intelligence (AI) techniques for the application area of robot communication. The study of the control and operation of multiple robots collaboratively toward a common goal is fast growing. Communication among members of a robot team and even including humans is becoming essential in many real-world applications. The survey focuses on the AI techniques for robot communication to enhance the communication capability of the multi-robot team, making more complex activities, taking an appreciated decision, taking coordinated action, and performing their tasks efficiently.

## I. Introduction

Artificial intelligence (AI) is being utilized to enhance sciences and technologies due to its amazing capability of dealing with big data, complexity, high accuracy, and speedy processing. Artificial neural network (ANN), fuzzy logic, neuro-fuzzy interference system (ANFIS), genetic algorithm, pattern recognition, clustering, machine learning (ML), particle swarm optimization (PSO), etc., are the familiar tools of AI, as depicted in Fig.1. AI has been employed in various areas such as engineering, science, medicine, computing, finance, economics and so on. Furthermore, it has been used to make machines smarter. Smart machine means the ability to make a machine perform intellectual tasks in an environment like or close to a human would do. Therefore, AI is part of the computer science and also arguably the most existing field in robotics. With a view to use AI approaches and robotics became an emerging technology at a rapid pace, offering many possibilities for automation tasks in many application areas such as domestic services, and space explorations [1], medical procedures, and military operations, collecting data about air pressure, temperature, climate, wind and so on [2]. Therefore, we can find functions of robots not only at work but also at home and industry, replacing many tasks that are dangerous and exhausting. Robots may be classified into two major types which are the service robots and field robots as shown in Fig.2.

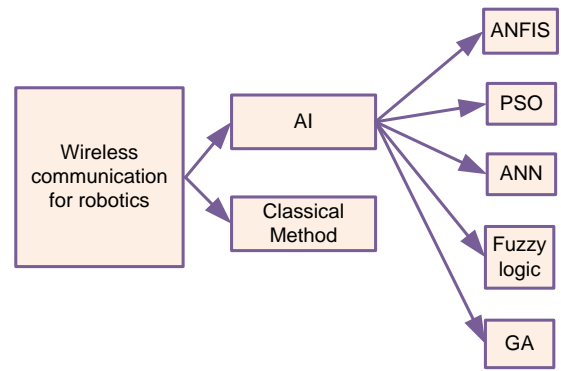

Fig.1 Wireless robotics communication using AI approaches

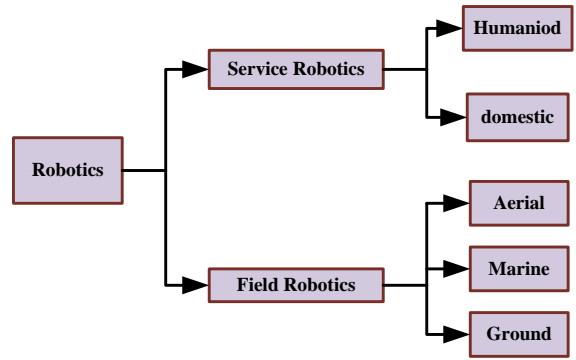

Fig.2 Robotics classification

Robotics has brought tremendous changes in various socio-economic aspects in our society. Robotic and AI aim is to create and understand machines capable of thinking and acting like humans. In view of this, robotics has the capability for self-learning [3, 4], self-organizing [5], self-reproduce[6, 7]. Nowadays, robots are becoming intelligent machines which use their artificial intelligence, abilities, and cleverness to perform tasks quickly and smartly. Robotics and AI together can make a machine do things in similar fashion as a human. In the future, robots will be everywhere, and they will help humans at any time in anywhere. To do that, communication among robots themselves and with humans is necessary. Therefore, robotics technologies have often combined with communication network technologies in both research and practice. The combination of both technologies will help robots to move autonomously and extend robotic functions to perform any given tasks effectively and efficiently.

Wireless network plays a vital rule in sharing and transmitting data between robots over the pervasive network, communicating with each other also with humans. A robot network is a group of robots that are communicated via wire or wireless communication technology performing tasks for a common goal in a coordinated manner. Communication can be teleoperated or autonomous [8]. Teleoperated robots refer to the robots which are fully controlled by human operators via communication networks. On the other hand, autonomous robots perform tasks or behave with a high degree of autonomy. Swarm robots are a group of robots placed in distribution fashion to perform a common task by sensing and sharing information via a communication network through internet or other means autonomously without direct intervention of a human. Efficient communication in cooperative robot team tasks is addressed in [9]. Communication among robots also needs to ensure collision prevention and maintain the communication link quality between them. Much research has been done in the area of robot-team communications [10, 11]. Intelligent robot communication is very important when the robots are either in the sky, on the ground, in water or in another environment, as shown in Fig.3.

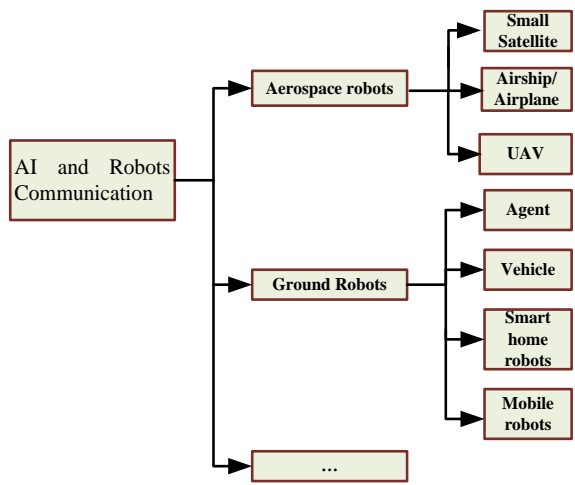

Fig.3 AI and robots Communication in different areas

Currently, AI and internet of things (IoT) make revaluation in the robotic domains applications. IoT represents the body part of robots that performs physical tasks, whereas the AI represents the brain part of robots that controls the physical robots according to the application needs [12, 13]. The difference between robotic things and IoT is that robots have intelligence concepts [14]. Pervasive robotics is closely interrelated with the internet of intelligent things (IoIT) [13]. The combination of robots, AI and IoT will result in robot systems having high capabilities to perform complex tasks autonomously. Therefore, with the help of IoT, robots can connect with each other and with humans easily, facilitating the high quality of exchanging data among them and with humans. Razafimandimby et al. [15] addressed the key technology for keeping connectivity between internet of robotic things (IoRT) to provide the desired quality of service (QoS) by using ANN.

This paper is focused on surveying AI methods for robot communication to enhance the communication capability of robots with each other, making activities, taking necessary and appreciated decisions, taking autonomous and coordinated actions and performed their tasks easily. The rest of the paper is organized as follows. Section II describes an overview of artificial intelligence and robotics communication. Section III presents intelligent aerospace robots communication, and intelligent swarm robotics communication is described in Section IV. Internet of robotic things is discussed in Section V. Future research directions are provided in Section VI and the paper is concluded in Section VII.

**II. Artificial Intelligence and Robotic Communication Overview**

Development of nano-antenna provides us with a new version of wireless communication called nano-wireless communication. Nano-wireless communication is very important for communication between small things. One of the applications of nano-communication is for communication between micro-robots [16, 17]. Always the communication between robotics uses ad-hoc network technologies because of independent nodes. There are robot communication protocols used for ensuring wireless connection between robots [18].

AI is difficult to define precisely. However, John McCarthy, known as the father of AI, defined AI as" the science and engineering making intelligent machines.". Therefore, AI refers to the ability of robotics systems to process information and produces outcome in a similar manner as humans do in learning, decision making, and solving problems. Furthermore, AI is growing continuously to make machines smarter and smarter. AI is used to make machines more intelligent, and machine intelligence is the ability of a machine to perform any intellectual task in any environment that a human would do, as shown in Fig.4. It is expected that robots serving humans or collaborating with humans must know the human needs with minimal communication from the humans. To this end, AI techniques were used to allow modeling and understanding of human intentions just by observing human physical movements [19, 20].

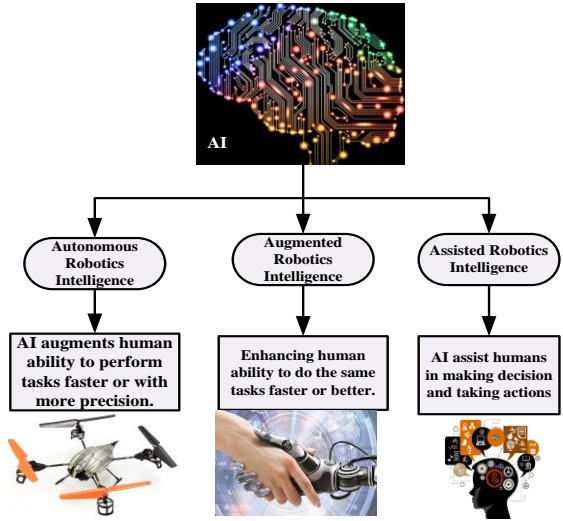

Fig.4 AI for Robotic Intelligence

Recently, ANNs have been applied to many diverse problems. ANN plays a vital rule in the environments that are nonlinear in nature such as communication and autonomy, as shown in Fig.5. The training of ANN is done using feeding the network input to learn. The network input transmits through the pattern with weights and neurons until it reaches an optimal. Training of a network could be done by supervised or unsupervised learning. The main advantages of ANN are the ability to be learned and trained from the failures as well as it can detect the relationship between nonlinear variables. Furthermore, ANNs output strongly provides accurate prediction and estimation from the input [21]. The statistical performance techniques and ANN for different applications are discussed [22]. Hence, the overview of using AI as a key technology to helping humans both on earth and in space makes this exploration feasible is shown in Fig.6.

ANN is a tool of artificial intelligence which is utilized for taking an appreciated hand-off decision when a mobile robot moves out of range of particular cell [23]. Furthermore, ANN was used for a mobile robot to communicate with its neighbor robots taking into account of their positions, orientations, and velocities in the same network coverage area [24]. Research has shown that using ANN helps reduction of the costs of communication and ensures the stability of the robots network. The reaction of robots to a human for a human-robot collaborative task is critical. Zhang et al. recently proposed an ANN-based method to predict human intention with which robots can take collaborative actions in time during the task [25]. Kelley et al. [26] also used learning technology to predict human intention. Furthermore, Li et al. [27] used ANN to coordinate an intelligent planner component of the hybrid agent for avoiding obstacles during moving. Using ANN helps a robot to avoid obstacles and keeps moving on.

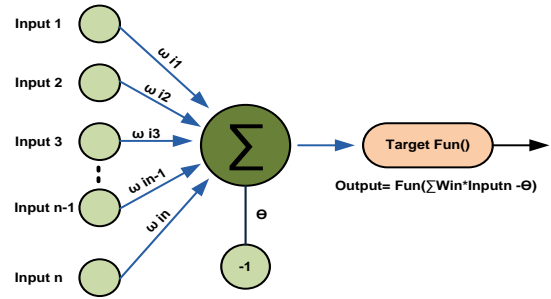

Fig.5 an ANN Network Structure

Robots are often used for search and rescue to access places and environments which a human cannot enter or safely enter. In this regard, communication is critical as rescuers need to send information and comments to robots as well as receive information from robots about the environment. Therefore, communication, robot, and AI indicate the safety and danger level for rescues, humans, and robots. Communication is also required for communication between robots and humans for accomplishing a task fast and easily. A robot navigation algorithm was demonstrated to be highly useful in guiding the robot to under any radio frequency identification (RFID) tag by a simple intelligent processing of the phase difference of the signal sent by the tag and received at both antennas of the RFID reader mounted on the robot [28].

An intelligent routing approach was proposed for a single unmanned aerial vehicle (UAV) which considers as delivering broadcasting unit [29]. Layer structure was divided based on hierarchical state and UAV operates in the upper layer for providing the broadcasting services to down layer such as a ground ad hoc wireless network. On the other hand, coverage area and network connectivity are the most important to design tradeoff between UAV teams. Therefore, Schleich et al. [30] developed and achieved an intelligence algorithm for UAV teams with network connectivity constraint. Development of performance protocols for an intelligent aerial vehicle was evaluated for providing an accurate aerial vehicles system [31]. Islam et al. [32] demonstrated that the proposed intelligent transmission control protocol (iTCP) exhibited a significant improvement in total network throughput and average energy consumption.

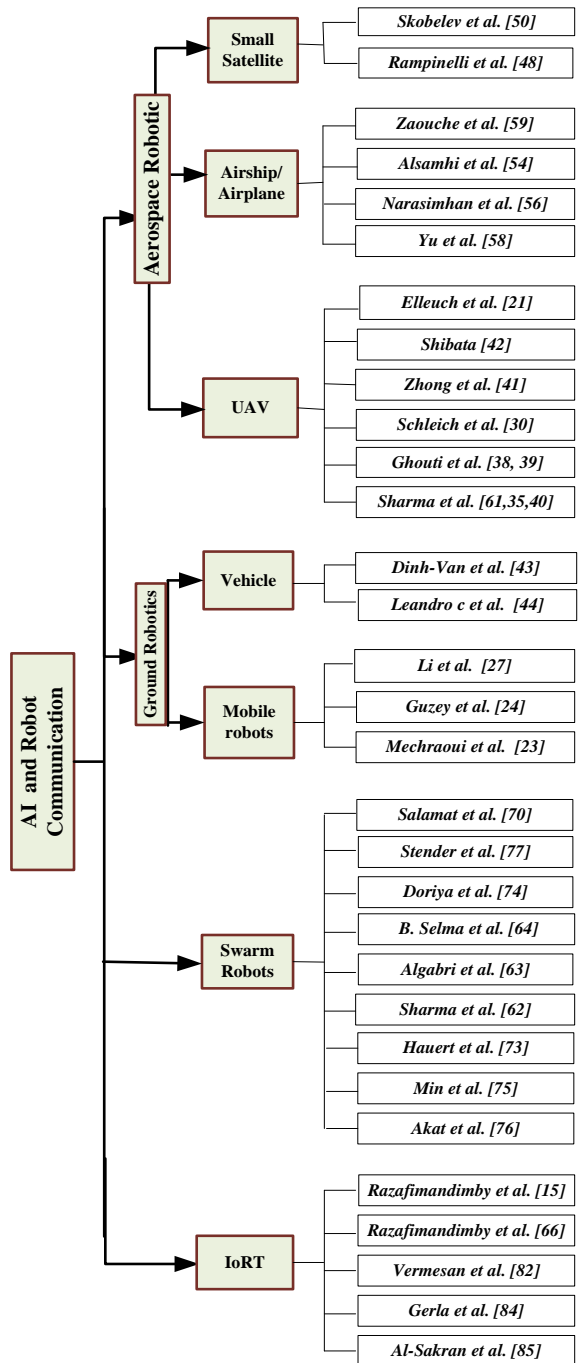

cooperation and communication between UAVs have been discussed and summarized using ad hoc networks [36, 37]. UAV was shown capable of identifying the subscriber and providing transmission of data between them [34]. The proposed UAVs were capable of delivering at a high data rate, effective bandwidth, and high QoS.

Elleuch et al. [21] predicted optimal paths for node mobility of wireless ad hoc networks by using ANFIS. Furthermore, Ghouti et al. [38, 39] predicted the same by using machine learning techniques. Ad hoc network technology was used for wireless communication between fly ad-hoc network (FANET) and ground networks [40]. Connectivity strategy between FANET and ground networks was coordinated by using ANN, fuzzy and genetic algorithm [35, 40].

To maintain the quality of wireless link and connectivity among robots, intelligence techniques are often required. Zhong et al. proposed an ANFIS technique which used a reinforcement learning method to maintain the desired coverage of wireless communication and control the quality of robots connectivity during performing their given tasks in an unknown environment [41]. On the other hand, Shibata proposed ANN which was trained using a reinforcement learning approach for negotiation between robots and end to end communication [42]. Dinh-Van et al. [43] improved an intelligence technique for indoor intelligent vehicle navigation to move autonomously using Wi-Fi fingerprinting and classification NN, which was capable of replace the GPS. The communication and security problems of intelligent robotic cars were investigated [44]. Table 1 shows the comparative of different involving approach for intelligent robots communication.

**III. Intelligent Aerospace Robotic Communication**

Intelligence space is targeted to construct an intelligence environment being able to monitor what is happening in the environment, and communicate with people. Intelligent space robots were built within a specific space using network distribution, sensors, actuators and processors [45]. The communication between the environment and functionalities of robots plays a key role to characterize the ubiquitous robotic space. The intelligent devices which are allocated in space were called distributed intelligent network devices (DINDs) [46]. These devices are used for environment observation. Communication between these devices DINDs enables the intelligent agents to understand

Fig.6 Overview of current studies

Ad hoc network technology was used for communication between robot teams [33, 34]. Coordinated robot teams on the ground and in the air have been investigated by [35]. The fuzzy logic approach was proposed to enhance signal transmission between UAV team and ground-robot team with the cooperation of an ad hoc network. Furthermore, the the events and perform tasks quickly and easily. Intelligent agents can monitor the environments and also deliver the information results to the users using communication technologies.

Table.1 Comparative analysis of different works involving approaches for intelligent robots communication

| Work | Proposed technique | Applications | Findings |
|------|--------------------|--------------|----------|
| *Razafimandimby et al., 2018[15]* | ANN | IoRT and desired QoS | They proposed intelligent technique to allow IoRT robots for working in the desired coverage area for provisioning desired QoS, and reducing energy consumption. |
| *Elleuch et a.l, 2015 [21]* | ANFIS | Mobility prediction | They proposed an effective technique for prediction the trajectory of an ad-hoc mobile. It was also demonstrated by testing on a time series prediction problems. |
| *Li et al., 2009 [27]* | ANN | Multi-robot system | They proposed intelligent approach for designing and coordinating hybrid agent framework effectively. |
| *Sharma et al., 2017 [35]* | Fuzzy | Fuzzy-bee colony optimization | The proposed optimization algorithm to provide efficient cognitive relaying between two ad hoc robots. |
| *Ghouti et al., 2013,2016[38,39]* | Machines learning | Mobility prediction | They proposed machines learning approach for mobility prediction. It was led to a significant improvement over conventional methods based on multilayer perceptions. |
| *Shibata., 2017 [42]* | ANN | Negotiation between robots and end to end communication | They established the communication learning by using initial weights in the sender agent. The robots were reached the goal. |
| *Girimonte et al., 2007 [49]* | AI | Intelligent space | They reviewed the applications of AI in the field of space engineering and space technology. |
| *Alsamhi et al., 2015 [54]* | ANN | Intelligent Hand-off for enhance QoS | The proposed ANN technique for enhancing hand-off rate and blocking rate. Hand-off decision between HAP and terrestrial system for enhance the QoS was greatly improved. |
| *Sharma et al., 2016 [61]* | ANN | UAVs assisted delay | They proposed intelligent approach for heterogeneous networks. Approach was capable of optimizing the allocation delays. It was not affect the capacity and coverage of networks. |
| *Selma et al., 2013[64]* | ANFIS | Navigation an unmanned vehicle | The proposed technique improved the control system adjusted in comparison to the ANN |
| *Algabri et al., 2017 [63]* | Fuzzy | Moving and tracking quadcopter | They proposed suitable technique for object-following tasks in surveillance applications such as crowd monitoring during Hajj. |
| *Razafimandimby et al., 2016[ 66]* | ANN | Desired QoS and global connectivity | They showed the efficient of IoRT robots using proposed approach, in terms of connectivity, convergence and energy consumption. |
| *Salamat et al., 2017 [70]* | PSO | Trajectory for Quadrotor UAV | They compered using different techniques in terms of execution time and effectiveness in finding the minimum length trajectory |
| *Min et al, 2016. [75]* | PSO and Genetic algorithm | Allocation of relay robots | They proposed PSO and genetic algorithm for finding the acceptable and efficient solution for relay robots allocation. |
| *Akat et al., 2010 [76]* | PSO | Search with a Multi-Robot System | They showed the performance of asynchronous PSO in a sense that the robots are able to move towards and aggregate in areas with high gas concentration around the maximum points of the gas concentration. |
| *Vermesan et al., 2017 [82]* | IoRT and AI | Concept of IoRT | They pointed the combination of Robotics, IoT and AI, which results in robots with higher capability to perform more complex tasks. |
| *Gerla et al., 2014 [84]* | Intelligent Vehicle | Internet of vehicles | They showed the evolution from intelligent vehicle grid to autonomous, internet-connected vehicles, and vehicular cloud. |
| *Al-Sakran et al, 2015 [85]* | Intelligent agent | IoT and Agent technology | They discussed the overview of monitoring the traffic based on IoT and mobile agent technology. |

The intelligent space agents move in space independently and monitor the environment using smart sensors as well as communicate with users on the ground. Intelligent space also described a place where many intelligent devices and sensors are distributed and communicated with each other using communication technologies [47]. Furthermore, Rampinelli et al. [48] described robot location and control in intelligent space using TCP/IP protocol for coordinating the communication.

Applications of AI play a critical role in the field of space engineering and space technology and identify[49]. Skobelev et al. [50] studied and presented a new intelligence technique for solving the problem of Earth sensing satellites by using multi-agent technology. Furthermore, Stottler described an AI techniques for automatic optimization, scheduling, and deconfliction of satellite communication [51]. However, the performance of loss and degradation of multi-spacecraft communication team was characterized optimally in [48]. The implementation of evolvable neural network control technology will enable a team of spacecraft to achieve formation flight with minimal supervision and sustain communication.

A high altitude platform (HAP) was considered as a complementary base station to a terrestrial mobile system, giving coverage in shadow zones [52-55]. HAP can provide back-up services to uncovered areas of the terrestrial systems, thus with the goodness of HAP, total capacity (QoS, in turn) in a service-limited area can be improved. A technique to recognize signal patterns of mobile subscribers using probabilistic neural network was introduced in the Rayleigh fading channel for enhancing QoS [56]. On the other hand, efficient hand-off algorithm enhances the capacity and QoS of cellular systems. Hand-off technique was used in wireless cellular systems to decide when and which base station should receive the particular call, without any service interruption [54]. Adaptive parameters such as user action speed, receive signal strength for pattern classification provide a multiple of criteria hand-off algorithm [57]. The neural network was trained to predict a user's transfer probabilities which represented the user movements [58]. ANN has been utilized to enhance hand-off techniques for auto-driving due to its ability to handle large data. Accordingly, a novel ANN based hand-off approach for efficient hand-off between high altitude platform and terrestrial system in a particular coverage area was proposed [54]. The performance of proposed technique reduced unnecessary hand-off and improved the hand-off rate. Zaouche et al. [59] provided an intelligence method for tracking target location of a network and video transmission using an aerial node.

Intelligent UAVs were not only used for providing large connectivity for a large area but used also for load traffic balancing [60]. Therefore, UAVs are capable of providing better reliability, capacity, and better connectivity in comparison to wireless communication networks. NN was used for dealing with network delay for UAVs cooperation in the next generation networks [61]. The effectiveness and accuracy of UAV optimal placement lead to reduce delays in UAV allocation. Efficient and accurate positions of the UAVs achieved larger coverage and delivered higher data rate. Furthermore, UAVs played a pivotal role for enhancing the coverage of the next generation of heterogeneous wireless network. Sharma et al. demonstrated how an intelligent UAV was used to enhance the 5G heterogonous wireless networks through enhance the capacity, throughput, SINR and reduce delay and error [62]. Algabri et al. proposed fuzzy optimization techniques to test and control UAVs during monitoring crowds participation in Hajj rituals [63]. However, B. Selma et al. proposed ANFIS control as an intelligent technique for navigate an unmanned vehicle [64]. These UAVs were not only used for monitoring but also processed to sending information to the center for taking reaction and appreciate decisions. UAVs can also detect the moving objects under their coverage area at different distance and elevation angle [49]. The newest version of an intelligent UAV system can facilitate the end user network nodes as well as mange the user searching, tracking and gathering [65].

ANN was proposed to provide the desired QoS and an efficient global connectivity of multiple mobile robots [66]. The author focused on an implementation of multiple IoRT and used ANN for maintaining the global connectivity and balancing the communication quality and network coverage.

The network architecture of an intelligent spy robot with wireless night vision camera was implemented in [67]. It was performed based on embedded systems and some appropriate software programs that can monitor the surroundings.

IV. Intelligent Swarm Robots Communication

Currently, machines are increasingly becoming smarter and able to gather data without human intervention. The reason behind this is to use artificial intelligence and machine learning technologies. Robotic communication with a human based on imitative learning, computing with words. Imitative learning is composed of model observation and model reproduction. ANN is applied for extracting spatial and temporal patterns of gestures and for utterance. Self-organizing map was used for clustering gestures. Experimental results showed that a robot could learn action patterns by incorporating some segments of human hand motions, and regulating the pattern-symbol relationship the cognitive environment through communication with the human [64]. Furthermore, Kubota et al. [68] discussed the benefits of ANN, self-organization map and genetic algorithm for the communication between partner robot and human.

Intelligent swarm UAV cooperative search strategies in a hazardous environment were presented and analyzed based on a number of deployed UAVs and search time [69]. PSO was used for multi-UAV trajectory optimization [70], and the genetic algorithm was simulated to make a comparison with the proposed algorithm regarding execution time and effectiveness in finding a minimum length trajectory. The UAV swarms were wireless ad hoc networks which form an aerial platform [71]. Formation of UAV swarm network among aerial platforms makes them suitable for many real applications domains such as military, civil monitoring, efficient tracking and searching areas, and weather monitoring. Surveillance using ad hoc swarm UAV network is dependent upon the data sharing, cooperation of taking appreciate decision. Failures in network will lead to decrease the performance of the network. Therefore, Sharma et al. [72] proposed an intelligent model called self-healing neural model to provide stability to all nodes in a network and provide necessary action for recovery of a node from a failed state to a stable state. Also, AI for a swarm UAV ad hoc network relaying was proposed with micro aerial vehicles [73].

One of the challenges of multi-robot communication is to control the movements of individual robots to perform their duties and reach their common goal safely. Doriya et al. [74] used PSO to coordinate multi-robot communication while cluster head gateway switch routing protocol was used forming clusters of robots. Genetic algorithm and PSO were used for finding an optimal distribution of robots for establishing an end-to-end wireless communication and relay robots in a disaster area [75]. The propagation signals, robot allocation, and robot path have been considered for a quick and long distance wireless communication in a disaster area. Asynchronous PSO based robotic search algorithm was tested in simulation environment and implemented with real robots [76]. Due to the inherent asynchronous operation of multi-robot systems, limited communication ranges as well as possible permanent or temporary agents or communication failures, the proposed algorithm is more suitable for multi-robot search applications compared to the other PSO variants. Stender et al. [77] studied a swarm of micro-robots collaborating to find a point of interest under noise and with limited communication in 2D space. Guided by a fitness function, the PSO algorithm is highly efficient to explore the solution space and find such an optimum.

## V. Intelligent of Internet of Robotic Things

The concept of IoT refers to things that may not be intelligent and does not include AI. However, IoRT is an intelligent concept which gives associated things, the ability of negotiation, reasoning, and delegation. Integrated of IoT and robots are growing upon ecosystem, whereas IoT devices, robots, and human communicate on fundamental cooperative. The target applications and technological implications of IoT- aided robotics were discussed [78]. Furthermore, Dutta et al. addressed the network security enhancement to IoT- aided robotics in the complex environment [79]. Dauphin et al. [80] reviewed the convergence in term of network protocols and embedded software for IoT robotic. The interaction of robotics and IoT have been investigated [81]. Therefore, AI, robots, and IoT will provide the next generation of IoT applications [82].

Razafimandimby et al. [66] implemented an intelligence technique to preserve global connectivity between IoRT robots by using ANN. ANN played a vital role in the balance between desired network coverage and desire QoS of communication of IoRT. IoIT and robot as a service (RaaS) interfacing within the cloud computing environment were discussed [83]. Architecture, interface, and behavior of RaaS defined, and the dependability design were the focus of the research. The internet of vehicles (IoV) has communications, storage, intelligence and learning capabilities to anticipate the customers' intentions [84]. Furthermore, agents provided an efficient mechanism for communication amongst networked heterogeneous devices within a traffic information system [85].

## VI. Future work

In the nearest future, all of things around us will be connected to internet computers, smart devices, mobiles and also huge variety of semi and fully autonomous robots. Therefore, IoIT goes further into transforming everyday objects into intelligent things, which communicate with each other and with humans. To do same, pervasive middleware was required for data transferring into actuators and also receiving data from sensors. Furthermore, Middleware can transfer AI processing data among terminals things and cloud according to the requirements. Here, the robots will be considered as middleware for integration of communication, data, for the internet of intelligent robotics things.

Few studies have been done in the area of communication between robots and intelligent techniques, but the topic is still very hot and should be taken into the consideration for its importance in many applications of our life and work. Keeping robots in the zone of network coverage is required for control, while control requires AI. QoS and quality of control will be the future work of intelligent robots communication. The combination of AI, ad hoc architecture and control infrastructure will be the perspective future in order to maintain the desired QoS and sufficient quality of control from multiple robots systems.

Formation of UAVs cloud with enhancement of the capacity is one of the challenges. UAVs can be used in huge formation to support huge number of users in large coverage area. However, using AI will help processing big data and deal with different

UAVs formation, protocols, and mobility in different environments.

The architecture of IoT and UAVs need to be specified in order to provide proper solution for enhancing the problems of management of UAVs in 5G technologies.

Nowadays, Power consumption became an importance issue, which should be considered for saving our environment green. In case of multi robots systems, the robots require for energy. Therefore, finding intelligence and efficient prediction techniques for reduced the energy usage will lead to green IoRT. Furthermore, prediction and intelligence technique will help also to predict routing tables, which would reduce the data exchange. Therefore, the life of the robots battery will be extended.

## VII. Conclusion

This article reviewed the intelligence techniques, which are currently being used to enhance robotic communication and making robots performance adaptive. Therefore, artificial intelligence is a key technology to help humans both on the Earth and in space, making exploration feasible. AI includes many approaches such as ANN, ANFIS, PSO, fuzzy logic, machine learning, and big data. The major advantage of using AI in robot communication is to coordinate individual robots' duties, protect the collision, and enhance the team performance, faster process and easier work. The intelligence added to the robots communication makes the robots smarter and perform tasks more effectively and efficiently. Furthermore, using AI as a technique for communication among robots and with humans via internet, lead to a new phase which is called IoRT. IoRT is a promising technology that makes intelligent things communicates with each other and with human.